# Superposing the Magnetic Spiral Structure of the Milky Way, on the stellar spiral arms – matching the unique galactic magnetic field reversal zone with two galactic spiral arm segments

Jacques P Vallée

Herzberg Astronomy and Astrophysics, National Research Council of Canada

ADDRESS   5071 West Saanich Road, Victoria, British Columbia, Canada V9E 2E7

ORCID *http://orcid.org/0000-0002-4833-4160*   EMAIL   jacques.p.vallee@gmail.com

**Abstract.**

To pinpoint the peak location  of the synchrotron total intensity emission in a spiral arm, we use a map of the spiral arm locations (from the observed arm tangent).  Thus In a typical spiral arm in Galactic Quadrant I, we find the peak of the synchrotron radiation to be located about  **220 ±40  pc** away from the inner arm edge (hot dust lane) inside the spiral arm. While most of the galactic disk has a clockwise largescale magnetic field, we make a statistical analysis to delimitate more precisely the smaller reverse annulus wiith a counterclockwise galactic magnetic field. We find an annulus width of  **2.1 ±0.3  kpc** (measured along the Galactic radius), located from 5.5 to 7.6 kpc from the Galactic Center). The annulus does not overlay with a single spiral arm  - it encompasses segments of two different spiral arms.

**Using a recent delineation of the position of spiral arms, the field-reversed annulus is seen to encompass the Crux-Centaurus arm (in Galactic Quadrant IV) and the Sagittarius arm (in Galactic Quadrant I).**

Thus the full Sagittarius-Carina arm is composed of : (1) a Sagittarius arm (in Galactic quadrant I) with a counterclockwise magnetic field, and (2) a Carina arm (in Galactic Quadrant IV). with a clockwise magnetic field.  Also the full Scutum-Crux-Centaurus arm  is composed of:  (1) a Scutum arm (in Galactic Quadrant I) with a clockwise magnetic field, and (2) a Crux-Centaurus arm (in Galactic Quadrant IV) with a  counterclockwise magnetic field.  Arm segments do not all have the same magnetic field direction.

For completeness, we display 6 known magnetised advancing supershells around the Sun (within 400 pc), pushing out the interstellar magnetic field.

## 1.  Introduction.

Recent advances in the locations of the spiral arms, partly through observing the locations of the *tangents* to spiral arms (see catalogs in Vallée 2022b; Vallée 2016a), have been done recently, showing  that different arm tracers were found separated from each other (dust,  masers, HII regions, broad diffuse CO gas).  We can thus compare them with the location of the total synchrotron intensity inside a spiral arm.

In the Milky Way, arm tangents as observed in different tracers have shown offsets from one another (see Vallée 2021; Vallée 2022a), with the dust at the inner arm edge, followed by masers, then young HII regions. Before reaching a spiral arm, the gas, clouds, stars all move at the circular orbital speed around the Galactic Center at about 233 km/s as observed (Drimmel & Poggio 2018). Upon reaching the inner arm edge, as predicted by the density wave theory (Roberts 1975), a shock front is produced, slowing down the incoming material. The shock/hot dust lane is physically visible, much like a 'road construction zone' which will slow all incoming cars inside. The shock and dust lane still moves slowly around the Galactic Center.  Thus, the gas, clouds, stars enter the spiral arm, and will later exit the spiral arm. One can thus talk of a 'relative speed with respect to the arm/dust lane ', which is the difference between 233 km/s and the spiral arm pattern (shock and dust)  speed mentioned.  One must realize that masers are short lived, so once formed through the inner arm (shock, dust), masers will become bright and be observable after a while, for a short time (transforming into something else).  Of course, new gas is still  entering the inner arm edge, and parts will become masers. So the 'maser lane' is continually renewed by new incoming gas. For a further review of the location / longitude of arm tracers, see Vallée (2017a).

**Arm tracers, arm pitch.** All arm tracers are going at the same speed, relative to the dust lane. That relative speed inside the spiral arm was evaluated at 81 km/s away from the dust lane (Vallée 2021). The masers and the HII regions are thus keeping their distances, relative to each others. Hence we observe small bright maser and micro-HII regions only when they are near the dust lane, but they continue along their circular orbit while they grow into something else. We observe large HII regions and stars later, when fully formed, thus not close to the dust lane. There is thus a correlation between age of a tracer, and an offset of that tracer from the dust lane. The Sun's Galactic radius is taken as 8.15 kpc (Abuter et al 2019), while the orbital circular velocity of the Local Standard of Rest is taken as 233 km/s. The velocimetry can thus be computed, for each spiral arm, at each distance from the Sun (Fig. 2 in Vallée 2008a).The pitch angle of a spiral arm can be measured globally. A global arm pitch angle would fit an arm over at least two Galactic quadrants (Fig. 1 in Vallée 2015; Table 1 in Vallée 2017b), employing a purely mathematical equation (equ. 10 in Vallée 2015) to get the global arm pitch angle from the galactic longitudes of the arm tangents in two Galactic quadrants; doing so yield a pitch value near 13º.

There is a published sketch of the oval orbital streamlines of the gas flow and old stars (in Roberts (1975)), moving clockwise around the Galactic Center. A weak spiral galactic magnetic field is known, going clockwise above a galactic radius near 7 kpc, and below a galactic radius near 5 kpc, and moving counter-clockwise in-between. While the stars align along a spiral shape in the galactic potential, so the magnetic field has a spiral shape, and the dust lane also has a spiral shape. Following the density wave theory (Lin & Shu 1964; Roberts 1975), **we adopt** a model where the circularly orbiting gas would bend a little when the gas enters a spiral arm following the shock lane at the inner edge of the spiral arm (below the co-rotation radius of the gas and wave).

**Magnetism origins, dynamos, pitch, locations.** Magnetic fields could be driven by stellar winds and supernova explosions. These magnetic fields could be further amplified by interstellar turbulence, by cosmic-ray buoyancy, by a turbulent galactic-wide magneto-hydro dynamo (MHD), and affected and sheared by the mean regular velocity field due to the 'differential rotation' $\Omega$ as a function of the galactic radius, as well as affected and enlarged by the mean helicity $\alpha$ of the interstellar turbulence. Hence this type of dynamo is called an $\alpha$ - $\Omega$ dynamo. A magnetic field pattern with a galactic density wave (driven by a dynamo) was predicted by Chiba & Tosa (1990).

The interstellar magnetic field has two main components: a regular (large-scale, coherently aligned) and a random (small-scale, turbulent) component, of about 2.5μ and 5μGauss, respectively, as seen from observed fluctuations in the Rotation Measures data, giving a total field strength near 6 μGauss in the disk. The random component is itself split into an isotropic random one (mid-scale, 3.5 μGauss) and an anisotropic random one (small-scale, 3.5 μGauss) – see Haverkorn (2015 – their Fig. 1). The regular magnetic field is often stronger in the interarms than in the arms (Shukurov 2002). Its energy density is comparable to that of other components in the Milky Way, making it dynamically important. Still, the total magnetic field strength, of about 6 microgauss, will not yield a 'magnetic control' of the gas. Cosmic-rays and relativistic electrons propagate through the disk along magnetic field lines.

In the Galactic disk, the observed large scale interstellar magnetic field has a spiral shape (clockwise or counterclockwise). Its pitch angle (deviation from a circular orbit around the Galactic Center) is measured to be slightly similar to the pitch of the stellar spiral arms (but not quite). The density wave theory predicted a slightly similar pitch (but not quite), allowing the lines of the magnetic field to slowly 'escape' the spiral arm. Thus the magnetic fields are not strongly 'attached' to spiral arms. Inside a spiral arm, the magnetic field lines have a slightly similar pitch angle than that of the spiral stellar arms, differing only by about 5º (van Eck et al 2015 – their Figure 10 and Section 5.2). Since the magnetic field lines are bent slightly than the gaseous arms, the magnetic field lines can thus exit the outer side of the spiral arm. Outside of a spiral arm, in the interarm regions, the pitch angle of the magnetic field lines may change with interarm location, thus may have a small pitch of 6º inward (Kronberg & Newton-McGee 2011), or a high pitch angle nearer 27º (Pelgrims et al 2021), or a pitch angle about 20º higher than the arm pitch (Van Eck et al 2015 – their Equa. 3),

Near the Galactic plane, within 5 kpc from the Sun, the magnetic field is symmetric (even) as one crosses from below to above the galactic disk. Towards the outer Galaxy (past the Sun in galactic radius), the disk magnetic field is nearly azimuthal and clockwise (Van Eck et al 2021 – their Section 4.1). Elsewhere in the inner disk, the direction of the magnetic field lines is spiral clockwise (as seen from the North Galactic pole), except for a region near the Sagittarius arm where it is spiralling counterclockwise (see below for details).

The magnetic field located very close to the Galactic Center (within 3 kpc) is difficult to measure (too few data yet). In addition, the origin of various '3-kpc features' emanating from the Galactic Center are still being debated – Vallée (2017c; 2016b) proposed that some of these 3-kpc features are the starting point of long spiral arms (Norma, Sagittarius). The magnetic field in the Galactic nucleus and its boxy bar are still unknown.

Most of the magnetic fields in the inner halo (close to the galactic disk) are taken as having the same direction below and above the Galactic plane. At 3 kpc above the disk, the total field strength is 2.5 µGauss (Haverkorn 2015). The vertical component of the magnetic field going into the halo is very small, perhaps negligible; there is no evidence for a vertical field in the Northern galactic hemisphere near the Sun, but perhaps there is a small component near 0.3 µGauss in the Southern hemisphere (Mao et al 2010). There is no evidence for a horizontal magnetic field component in the halo (Mao et al 2010 – their Section 7). Some have sought a magnetic field in the sub-halo (with a vertical distance between 0.5 kpc and 2.5 kpc) and found ambiguous results; this region is affected by exploding interstellar bubbles and Loops (Xu & Han 2019).

Does a magnetic field line follow the spiral arm shape in Galactic radius, toward the Galactic Center (Fig. 11 in Van Eck et al 2011) – and then up in the Galactic halo ? Or is it going back from the halo into the disk through another spiral arm?

Using the galactic longitudes of tracers of spiral arms, in **Section 2 we** find the location of the peak intensity of the total synchrotron emission, inside an arm width, away from the dust lane at the inner arm edge. In **Section 3 we** find the exact boundaries (top and bottom in galactic radii) of the counterclockwise magnetic field areas, and we superpose these magnetic areas with precise stellar spiral arms. In **Section 4** we assess the known local turbulence from supershells near the Sun, on the regular magnetic field. Do turbulences cover the whole Galactic disk ? In **Section 5** we conclude.

## 2. Location of the total synchrotron emission peak, among the various tracers of a spiral arm

Recent telescope scannings in Galactic longitude have shown a separation of different arm tracers, inside each arm width. Where is the Dust, low-density and high-density thermal gas, synchrotron electrons, HII regions, young and old stars are all found in spiral arms, but all tracers do not coincide exactly as each one is displaced by up to 300pc from each other, as known since the work of Vallée (2014). This Section determines where the strongest magnetic field lies in an arm. Earlier scans in Galactic longitudes, along the disk of the Milky Way, at radio wavelengths gave results on peak intensities (Table 3 in Vallée 2016b). At 408 MHz, the peak intensity is at galactic longitude 310$^\circ$ in the Crux-Centaurus arm, then at longitude 328$^\circ$ in the Norma arm, then at 339$^\circ$ in the Perseus start arm, then at 032$^\circ$ in the Scutum arm, and at 049$^\circ$ in the Sagittarius arm. In each case, this radio total intensity peak is located inside the arm (past the near-IR dust peak, but before the broad diffuse CO gas peak).

**Figure 1** shows the position in an arm width, where each arm tracer peaks in a different specific galactic longitude. Hence overlaying the Scutum arm over the Sagittarius arm and over the Norma arm (in Galactic quadrant I) allows the mean location of different arm tracers to be seen (x-axis).The x-axis is the azimuthal angle as converted into a linear separation (from the proper distance from the Sun to the spiral arm). The arm tracer are observed to be separated, with at left a red zone (hot dust NIR and MIR tracers), then the orange zone (radio masers, Fir [CII} and [NII] lines, colder FIR dust), then a green zone (HII regions radio recombination lines, thermal free electrons, synchrotron peak intensity), and lastly a blue zone (brad diffuse CO emission peak intensity, NIR intensity old stars). The orbiting gas is from left to right, on the x-axis.

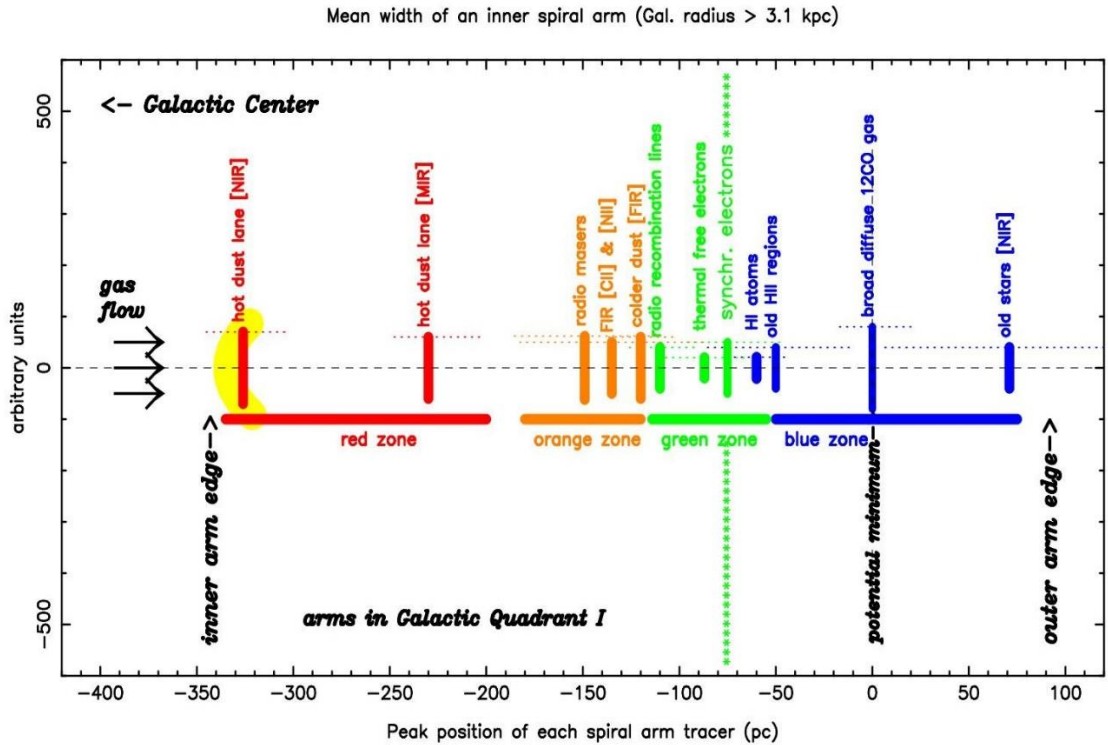

*Figure 1. A view of a spiral arm in Galactic Quadrant I, including and averaging over the Norma arm (near l=18°), the Scutum arm (near l=30°), and the Sagittarius arm (near l=50°). In a spiral arm seen tangentially, the arm's cross-section shows a separation between each arm tracer (x-axis), from the dust lane red zone, at left) at around -300pc, to the orange zone (radio masers, young stars) at around -150pc, to the green zone at around -80pc (the synchrotron electron total intensity peaks at a wavelength of 74cm is shown with a vertical green line of asterisks, near -80 pc), and to the blue zone (broad diffuse CO line peak, old stars) at around 0 pc. Thus the offset from the hot dust lane to the synchrotron emission peak is about 220 pc.*

As can be seen in this Figure, in Galactic Quadrant I, the peak of the synchrotron radiation (total intensity) is located near -80 pc, away from the inner arm edge (hot dust lane at -300 pc), thus a separation of about **220 ±40 pc** from the hot dust lane. This would be the central peak position of the otherwise wide extent where the magnetic field is located, inside the spiral arm. This peak synchrotron location would indicate the location of the total (regular + turbulent) magnetic field strength, with a predominance for the turbulent field. The polarised intensity (regular magnetic field) is known to be concentrated elsewhere, in the interarm (in-between two spiral arms). Original data (catalogs of arm tracers were published in Vallée (2016a), Vallée (2022b).

Averaging over several arms in the same Galactic quadrant will give a mean typical picture of these arms, as the same arm morphology is expected for an isolated galaxy.

Similarly, in Galactic Quadrant IV, the peak of the synchrotron radiation is located near 85 pc, away from the inner arm edge (hot dust lane at 370 pc), thus a separation of about **285 ±30 pc** from the hot dust lane (Vallée, 2022b).

For the Sagittarius arm, the synchrotron emission peak is located at Galactic longitude l= 049° **±1°** (Beuermann et al 1985). This value is the same within the error as the longitude of the peak RM longitude of Shanahan et al (2019) inside this arm.

*We expect that the exact position of the peak total synchrotron emission within a spiral arm, here seen between the masers and starformation region (orange lane in Fig.1), and the broad diffuse CO lane and Potential Minimum of the wave (blue lane in Fig.1), should be a future **prediction** of the density-wave theory combined with of a Magneto-Hydro-Dynamo theory.*

### 3. New results: Matching the single field reversal zone with the spiral arm positions.

Previous mapping of the interstellar magnetic field direction in the galactic disk yielded this picture. To measure the direction of the interstellar magnetic field, we can measure its effect on electromagnetic waves going through the interstellar medium. We can use point sources or extended sources, at various galactic longitudes and latitudes.

Externally, synchrotron polarized radio waves from distant point sources (quasars and distant galaxies), detected with a polarimeter, allow the mapping of the magnetic field in the Milky Way through Faraday rotation. A polarised radio wave rotates its polarisation angle, as it travels, by a measurable amount (the Faraday Rotation Measure). The amount of rotation is proportional to the distance travelled, to the magnetic field strength, to the thermal electron density, and to the square of the wavelength– see Vallée (2017a). The rotation measure (RM) of many quasars and galaxies can be measured through a spiral arm, to derive the Faraday rotation of the radio wave. Haverkorn (2015) provided a recent observational review on this topic, including the basic terminology employed.

Internally, in the Milky Way, synchrotron polarized light coming from relativistic electrons, trapped by the galactic magnetic field lines, will have their electric vector perpendicular to the interstellar medium's magnetic field line. Thus magnetic direction and strength can be measured at various galactic longitudes and latitudes. Also, within dust aligned by an ambient magnetic field, a polarised incoming radio wave will go through the dust (along the line of sight) if the wave's polarisation is perpendicular to the magnetic field, but the radio wave will be scattered (across the line of sight) if polarised otherwise.

**Observed radial Region with a magnetic Field reversal.** Most published models (fitted to the observations) have a single annulus region with a magnetic field reversal – a field reversal should be added to a model, only when proven necessary by the observational data – see discussion below.

Using a Faraday Rotation Measure process, the magnetic field direction is found; it is not azimuthal or circular; but in a spiral shape (similar but not identical to the stellar spiral arms).

The first magnetic field reversal in galactic radius was found in 1980 (Section 8 in Simard-Normandin & Kronberg 1980; Fig.5 in Thompson & Nelson 1980; see also Vallée 2011).

The second magnetic field reversal in galactic radius was found closer to 1990. Prior to 1990, it was difficult to observe enough pulsars Rotation Measures far from the Sun toward the Galactic Center (down to a maximum of 2 kpc from the Sun, see Fig. 4 in Rand & Kulkarni 1989).

Together, the first reversal and the second reversal constitute an annulus, within which the counterclockwise magnetic field was found.

Many different values were employed for the distance between the Sun and the Galactic Center, many taking 8.5 kpc as such. We collated each report on this annulus from the literature, and we re-scaled them all to the current value of 8.15 kpc for the solar distance to the Galactic Center, then we performed simple statistical means on the magnetic field reversal zone (bottom, top from the GalacticCenter).

Thus Vallée (1991 – his Fig.1) showed a clockwise magnetic field in the galactic disk, except between an annulus from a Galactic radius where the field is counterclockwise. This annulus is at 5.7 kpc to 7.5 kpc, for $R_{Sun}$ = 8 kpc, corresponding to an annulus from 5.8 to 7.6 kpc for the latest $R_{Sun}$ = 8.15 kpc.

Also, Rand & Lyne (1994 – their Fig.5) showed a magnetic field reversal in an annulus starting at 8.1 kpc from the Galactic Center and ending at 3.0 kpc from the Sun, using a 8.5-kpc solar distance from the Galactic Center. Their annulus is at 5.3 kpc to 7.8 kpc for $R_{Sun}$ = 8.15 kpc.

**Regions of enhanced thermal electron density (spiral arms).** In the absence of spiral arms, many employed well-used models for the distribution of the thermal electron density in the Galactic disk: the TC1993 model (Fig.4 in Taylor & Cordes 1993), the NE2001 model (Cordes & Lazio 2002; Fig.9 in Cordes & Lazio 2003), the YMW2016 model (117 parameters – Fig. 9 in Yao, Manchester & Wang 2017) have served in the past, within their uncertainties. They basically followed 4 long spiral arms, as determined from their effects on the line of sight to pulsars and other tracers. The first two models employed a Sun-Galactic Center distance of 8.5 kpc, while the last one used 8.3 kpc.

***Recent arm model, based on the broad diffuse arm tangents near the Potential Minimum of the wave.*** *Here we propose* to employ the many properly fitted arm tracers (not just the electron density) to delineate the spiral arms. Thus we propose that a simple model of the free electron distribution should follow the better known locations of many tracers of spiral arms, based on the precise observations of the tangents to the spiral arms as seen from the Sun, and the fact that most arms contain copious amounts of free thermal electrons (Vallée 2021).

The existence of 4 long spiral arms, with a log-normal shape, is known since the work of Georgelin & Georgelin (1976). We used a Sun-GC distance of 8.15 kpc, a log-spiral arm shape, arm pitch of 13.4° (Fig. 4 in Vallée 2022a),  each arm starting at 2.2 kpc from the Galactic Center (Vallée 2016b). This Galactic-wide model is basic; here we are not interested in fitting local deviations or small-scale structures (supernova explosion, HII region bubble, Local Arm(let), and clumps or voids). The fit is quite good (Fig. 3 in Vallée 2022a).

See **Table 1**, giving *a statistical* averaging to get a more preciset location of the Counterclockwise annulus*, after a radial re-scaling to a common position of the Sun's galactic radius.*

**Table** 1 collects these independent data from the literature. We can do here an averaging of the annulus's bottom Galactic radius, and another averaging of the annulus's top Galactic radius.   All the radii above for the top of the annulus are at or above 7.1 kpc  (except for the Brown et al 2007 paper and the Andreasyan et al 2020 paper). Including those papers, the root mean square (rms) is large (±0.20 kpc). Excluding these two papers, the rms decreases by a factor of two, and  we get a mean bottom annulus radius at 5.5 ±0.13 kpc and the mean top annulus radius at 7.6 ±0.10 kpc  - thus covering a width of **2.1 ±0.3 kpc** along the Galactic Meridian  (the line joining the Sun to the Galactic Center).

=================================================================================

Table 1. Boundaries (top, bottom) of the Counterclockwise magnetic field, in an annulus

| Top Galactic radius [a] kpc | Bottom Galactic radius [a] kpc | Reference | Old Distance to Gal. Center kpc |
|---|---|---|---|
| 7.6 | 5.8 | **Vallée (1991 – Fig.1)** | 8.0 |
| 7.8 | 5.3 | **Rand & Lyne (1994 – Fig.5)** | 8.5 |
| 7.5 | 4.8 | **Indrani & Deshpande (1999 – Fig.4)** | 8.5 |
| 7.4 | 5.1 | **Vallée (2005 – Fig.4)** | 7.2 |
| 5.9 | 4.2 | **Brown et al (2007 – Fig.4)** | 8.3 |
| 7.2 | 5.8 | **Sun et al (2008 – Fig.10a)** | 8.5 |
| 7.7 | 5.8 | **Noutsos et al (2008 – Fig.7)** | 8.5 |
| 8.0 | 5.6 | **Van Eck et al (2011 – Fig.6)** | 8.5 |
| 8.0 | 5.6 | **Xu & Han (2019 – Fig.4)** | 8.3 |
| 7.0 | 4.6 | **Andreasyan et al (2020 - abtractt)** | 8.5 |
| | | | |
| 7.41 ±0.20 | 5.26 ±0.18 | **Mean and rms (all data)** | |
| 7.65 ±0.10 | 5.48 ±0.13 | **Mean and rms (excluding the data from Brown et al. 2007 and from Andreasyan et al 2020)** | |
| 7.58 ±0.21 | 5.48 ±0.22 | **Mean and rms (all data recently published – since 2008 or later)** | |

**(a):** Galactic radius values, after re-scaling here to a common $R_{Sun}$  value of 8.15 kpc.

=================================================================================

**Figure 2** shows the current fitted model to the arm tangents, as well as the annulus with a counterclockwise magnetic field direction.

From the Earth, we find the magnetic field reversal (Table 1) to occur in Galactic Quadrant I  between galactic longitudes l =46° to l=71°, encompassing only the Sagittarius spiral arm; see Fig. 2 in Ordog et al (2017) showing a jump in RM data in that longitude range.  Interestingly, this magnetic field reversal thus excludes the Scutum spiral arm.

 From the Earth, we find the magnetic field  reversal to occur in Galactic Quadrant IV  between galactic longitudes l =289° to l=314°, encompassing only the Crux-Centaurus spiral arm; see Fig.9(j) in Sun et al (2008) and Fig. 1 in Vallée (2008b), showing a jump in RM data in that longitude range.  Interestingly, this magnetic field reversal thus excludes the Carina spiral arm.

Hence the long Sagittarius-Carina arm is separated in a counterclockwise magnetic field in Sagittarius (in Galactic quadrant I) and a clockwise magnetic field in Carina (in Galactic Quadrant IV). Likewise, the long Scutum-Crux-Centaurus arm is separated in a clockwise magnetic field in Scutum (in Galactic quadrant I) and a counterclockwise magnetic field in Crux-Centaurus (in Galactic Quadrant IV). These separations of the magnetic field direction (clock or anticlock) depend on the Galactic radius, thus showing the magnetic field **crossing of the spiral** arm when needed for the dynamo action.

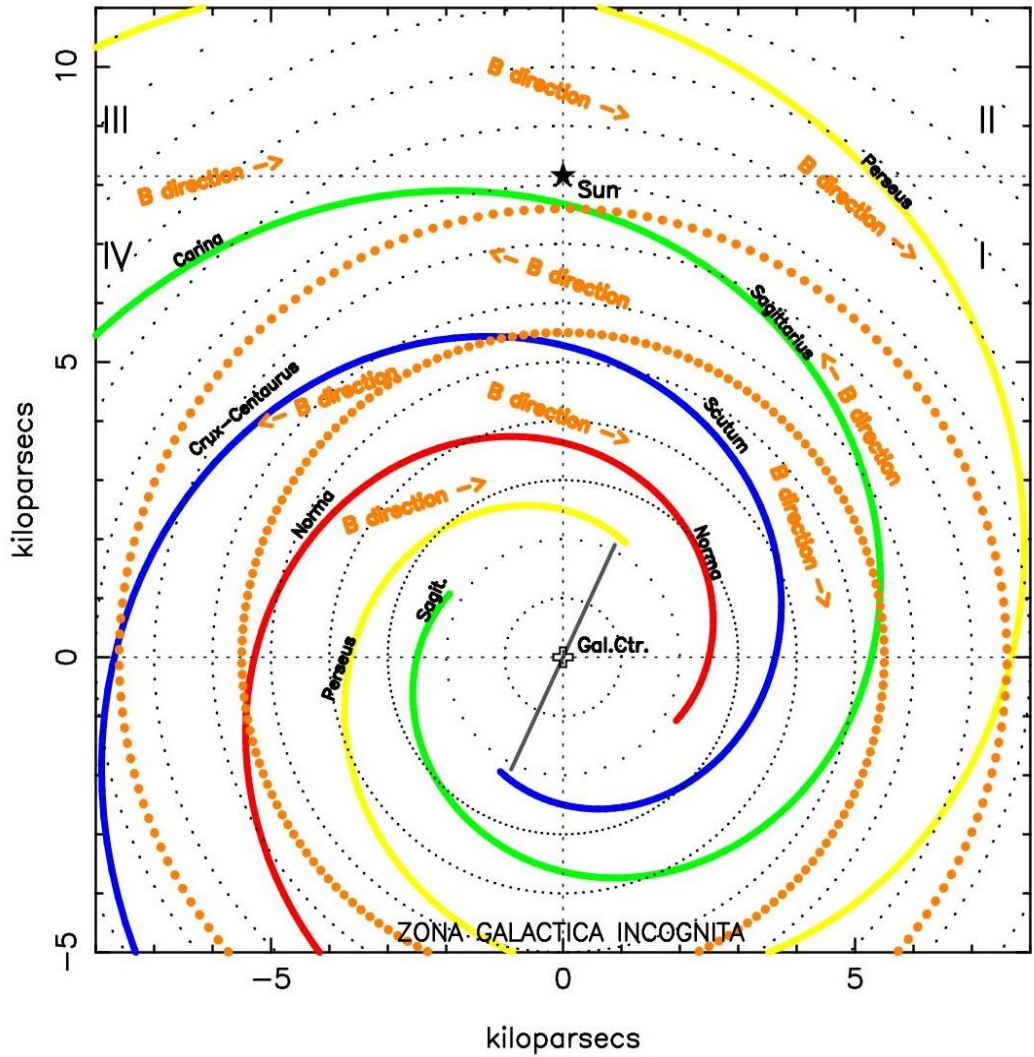

*Figure 2. The large orange dots forming an annulus, one at 5.5 and one at 7.6 kpc, show the boundaries of the annulus with a counterclockwise magnetic field; its magnetic field direction (B dir) is indicated with an orange arrow, located south of the Sun. Outside of the annulus, the clockwise magnetic field directions are shown (orange arrows, north of the Sun, and south of 5 kpc). For this rendering, the arm pitch angle is at -13.4º, and each arm starts at 2.2 kpc from the Galactic Center. The Sun is at 8.15 kpc above the Galactic Center. The largely unknown area below the Galactic Center is often labeled 'Zona Galactica Incognita'. Galactic quadrants I to IV are shown. Arms are adapted from Vallée (2021).*

Looking at the map (Fig.2), the counterclockwise annulus essentially covers (1) the Sagittarius arm in Galactic quadrant I, and (2) the Crux-Centaurus arm, along with adjacent portions of its nearby interarms. No change is seen when crossing the interarm – arm border.

**Arm tangents.** Does the magnetic field get amplified in a spiral arm? It appears so.

Earlier, Vallée et al (1988b – their fig.1) found an excess RM of -75 rad/m$^2$, coming from the large *Scutum* spiral arm (from 25º to 40º of Galactic longitude), with a clockwise direction as seen from the North Galactic pole.

Reissl et al (2020 – their fig. 5) made a magneto-hydro-dynamical simulation of the galactic disk that includes the four spiral arms, and predicted an excess Faraday rotation measure (their fig.5) near the longitude of the tangent to each spiral arm.

Their model used a mean offset of about 4º between the arm tangent galactic longitude of the CO peak and the arm tangent galactic longitude of the dust peak in the same arm (their fig. 6 and their table 1). Observationally, Vallée (2016a – his table 1) measured a mean offset of 3.2º ±0.3º.

Shanahan et al (2019) measured the Sagittarius arm (from 39º to 52º of galactic longitude; in averaging steps of 0.2º) close to the **arm tangent** as seen from the Sun, and they found an excess Faraday rotation measure (+3000 rad/m²) inside the *Sagittarius* spiral arm, with RM peaking at longitude near 48.2º (see their Figure 3a).

Ma et al (2020) found an excess rotation measure distribution, localised near the **tangent** to the *Sagittarius* spiral arm (longitude from 20º to 52º), with a large averaging step of 5º, visible from l=41º to 55º and peaking near l=48º (RM= +400 rad/m²; their Fig. 13) – similar to the results of Shanahan et al (2019) who used a smaller averaging bin.

Ma et al (2020 – their Section 5.6) looked at the magnetic field between Galactic longitudes 20º and 52º, and improved on the model of Van Eck et al (2011) by adding a localised even-odd vertical parity field, associated with the tangent to the *Sagittarius* arm at a distance of about 5 kpc from the Sun.

      **Unproven claim of multiple field reversals at various galactic radii, in the disk.** A single group has proposed alternating magnetic field directions (clockwise in each arm; counterclockwise in each interarm) – see Fig. 12 and 13 in Han et al (2006) or Fig.8 in Han (2017). There are problems with trying to observe this alternating arm-interarm model (not enough observational data), and its complexity to create it with a dynamo - a 4-arm spiral model will then have 4 magnetic field reversals in Galactic quadrant I, and 4 magnetic field reversals in Galactic quadrant IV; also, the galactic longitudes of their arm tangents from the Sun are mostly assumed, not observed (and not defining well the start and the end of each magnetic field reversal region); hence large galactic longitudes have little observational data (not feasible to test this model). As mentioned above, the known observational data (Table 1) favor a single region with a magnetic field reversal over roughly 2.1 kpc in radial range.

      **Predicted radial region, with a magnetic field reversal (dynamo theory).** Such a 'localised magnetic field reversal' had been predicted in some 'dynamo' theories - see an early review in Vallée (2008b). Only some dynamo models and some initial conditions can generate a magnetic field reversal, like those in our Milky Way galaxy. Here are some examples.

      Ruzmaikin et al (1985 – their Fig.3 and Fig.5) computed the radial potential and the radial modes (y-axis) of the galactic dynamo, along the galactic radius (x-axis). These were adapted to the Milky Way in Vallée (1991 – his Fig.3), showing a counterclockwise magnetic field (annulus) in the Sagittarius spiral arm, and a clockwise magnetic field elsewhere in the disk.

      Poezd et al (1993 – their Fig. 4a) could also predict a galactic dynamo with two radial reversals.

      Moss et al (2012 – their Fig. 4f) obtained a region with a magnetic field reversal by continuous injection of turbulence (from supernova explosion) and a rapid galaxy rotation (large dynamo number).

      Moss & Sokoloff (2013 – their Fig.2) employed a dynamo model, and various initial conditions, to get a large-scale magnetic field reversal below the Sun's location.

      Dobbs et al (2016 – their Fig.4) used smoothed-particle magneto-hydro-dynamical simulations, using an imposed spiral potential, and have predicted a magnetic field reversal between a galactic radius of 4 to 6 kpc..

      Shukurov et al (2019 – their Fig. 7) employed a mean-field dynamo with turbulent transport (alpha-effect and beta diffusion), and some input parameters (their Table 1), producing a reversed magnetic field annulus (between 7 and 12 kpc from the Galactic Center).

      Mikhailov & Khasaeva (2019 – their Fig. 4) employed the mean-field dynamo, some specific initial conditions, and obtained one magnetic field reversal between 1 and 5 kpc from the Galactic Center, with the reversal able to move at a speed of 1 km/s (their Fig. 7), giving 1 kpc in 1000 Myrs.

      Andreasyan et al (2020 – their Fig.5) proposed a thin disk dynamo approximation (very small magnetic field in the vertical z-coordinate), yielding a counterclockwise annulus below the Sun's location in the disk after a certain time.

      This arm model, fitted to arm tracers (Figure 2), and the radial location of the reversed annulus (Table 1), led us to identify the observational location of the annulus with respect to the nearby spiral arms. *We expect that existing theories (Galactic dynamo, MHD) should **predict** these results, annulus size, top and bottom Galactic radius, and including the consequences of a long spiral arm encountering a magnetic field reversal.*

      **4. Turbulence, local and everywhere.**

There is a turbulent magnetic field, of about the same strength as the strength of the regular magnetic field, or more. The turbulence can be made by various physical phenomena: interarm islands, Local Arm(let), a supershell around an OB star association, or the ejecta of a supernova.

**Local Arm and interarm islands.** The Sun is located in an interarm, in between two long spiral arms (Sagittarius arm and Perseus arm). Other small interarm islands are known within our Galactic disk (Vallée 2018; Vallée 2020). The magnetic field strength there is not known (too few data). The turbulent magnetic field seems to thrive around localised stellar associations.

**Local supershells.** The presence of large interstellar supershells (going into the halo) near the Sun, each one centered on an OB star association, can cause the interstellar magnetic field to loop around them (Fig. 10 in Vallée 1973; Fig. 2 in Vallée & Kronberg 1973; Fig. 1 in Vallée 1984). Since these shells expand into the halo, they would distort a nearby halo field there and compress the magnetic field in the local interstellar medium; hence the local supershell's contributions to the halo pulsar rotation measures are significant (Xu & Han 2019).

Other smaller (less than or about 2º) nearby localised anomalies are seen in extensive RM maps. Van Eck et al (2021 – their Section 4.2 and Figure 6) discussed the effect of major HII regions, such as the Sh2-230 complex and the W3/W4/W5 complex.

**Table 2** here shows some data for the large nearby magnetic supershells – there could also be many more distant ones. Using 9 nearby supershells, it was found earlier that [the shell gas density] goes inversely as [the shell thickness $\Delta r$ / the shell radius r], and also that the shell magnetic field strength $B_s$ goes inversely as [the relative shell width $\Delta r$ /r]; thus the smaller the relative thickness, the more compressed the shell gas and the shell magnetic field were observed (Fig. 1 and 2 in Vallée 1993a). In addition, it was found that the shell magnetic field $B_s$ and the shell gas density $n_s$ were related, as $B_s$ is being directly proportional to $n_s$ (Fig. 1 in Vallée 1993b). In the nearby cool supershells, $B_s$ has a range from 2 to 10 µGauss, while $n_s$ ranges from 0.04 to 2.5 cm$^{-3}$ (Table 2 in Broten et al 1985).

==============================================================================

Table 2. Nearby large supershells, within 400 pc of the Sun, with an angular diameter larger than 30º and a known magnetic field (a) (b)

| Supershell name | Center long., latit. | Solar Distance | Shell diameter | Shell thickness | Shell magnetic field | Model field (using rotation measure) |
|---|---|---|---|---|---|---|
| Loop II – Cetus arc | 110º, -32º | 100 pc | 100º | 5º | 17 µG (a) | Fig. 11 in Thomson et al (2021). |
| Fan | 135º, +0º | 100 pc | 40º | - | 24 µG (b) | Fig. 6 in West et al (2021) |
| Eridanus shell | 195º, -30º | 400 pc | 38º | 3º | 8 µG  (a) | Fig. 1 in Vallée et al (1988a) |
| Monogem ring | 203º, +11º | 300 pc | 30º | 5º | 4 µG (a) | Fig. 2 in Vallée et al (1984) |
| Gum nebula | 260º, +0º | 400 pc | 36º | 4º | 2 µG (a) | Fig. 3 in Vallée & Bignell (1983) |
| Loop I – North Polar Spur | 329º, +18º | 130 pc (c) | 116º | 8º | 6 µG (a) | Fig. 2 in Vallée & Kronberg (1973) |

(a): Reference for coordinates, distance, shell diameter, shell thickness, shell magnetic field:  Tables 1 and 2 in Vallée (1993a)
(b): Reference for coordinates, distance, diameter, magnetic field: Sections 2.2 and  4.1 in West et al (2021).
(c): Solar distance from Panopoulou et al (2021) was revised, with parts ranging from 105 to 135 pc.

==============================================================================

**Figure 3** shows an exploded version, around the Sun, adding  the magnetised  supershells listed in Table We expect these expanding supershells to bend the large-scale galactic magnetic field, around each supershell. One could imagine the large-scale field, going from left (longitude 270º) to right (longitude 90º), being bent locally around each expanding supershell, using radio data – see Figure 10 in Vallée (1973),  Figure 2 in Vallée & Kronberg (1973) and Figure 1 in Broten et al (1985). This was confirmed later optically, with the local curvature of the magnetic field as derived from starlight polarisation (Heiles 1996).

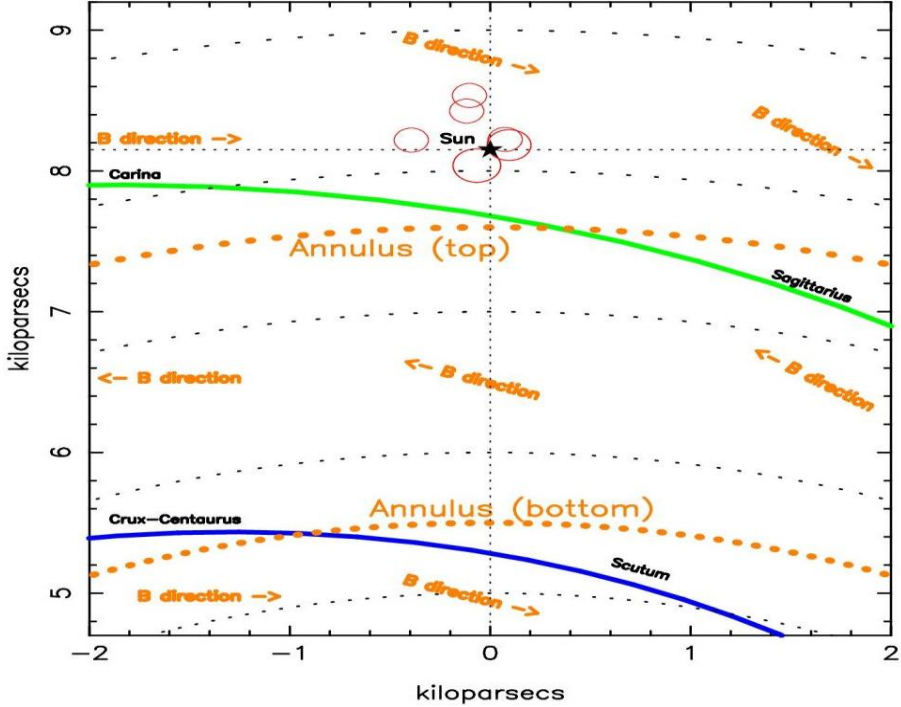

*Figure 3. A map near the Sun, with the Sun shown as an asterisk at 8.15 kpc from the Galactic Center. We added the 6 nearby magnetic supershells (in red), as listed in Table 2, all with a centroid within 400 pc of the Sun. The orange annulus with the reversed (counterclockwise) magnetic field is shown. Two nearby spiral arms (blue Scutum-Crux-Centaurus arm, green Sagittarius-Carina arm) are shown.*

There is no known direct link between a type of turbulence (supershells, Table 2) and the galactic zone of reverse magnetic field direction (Table 1). Here we can point what the observations showed. Those magnetic supershells in Figure 3 are all close to the Sun and inside a clockwise galactic magnetic direction, but we expect many more supershells everywhere (even inside the counterclockwise galactic field annulus in Figure 3 (when we reach better observational sensitivities).

A map of Rotation Measure observed in Galactic Quadrant I, with Galactic longitudes 54° to 70° and Galactic latitudes -3° to +4°, showed a separation of positive RM from negative RM (see the diagonal dotted line in Fig. 2 in Ordog et al 2017; see also Fig.6 in Van Eck et al 2021); however, they attributed the short diagonal dotted line to changes in the large-scale Galactic-wide polarised synchrotron emission, linked to the known Galactic magnetic field reversal between us and the Sagittarius arm at about 2 kpc from the Sun in that direction

**Here we propose** that the nearby stellar association/supershell known as Loop II (see the arcing dotted line in Fig.1 of Uyaniker et al 2001, near longitudes 64° to 70° and latitudes -5° to +4°) seems to follow very well the short diagonal dotted line in Fig. 2 of Ordog et al (2017) and Fig. 6 in Van Eck et al (2021) separating positive RM versus negative RM. A recent model of Loop II was published by Thompson et al (2021 – their Fig.11), at about 0.2 kpc from the Sun in that direction.

What is the role of the interstellar turbulence, with respect to the Galactic-wide magnetic field, and with respect to the location of the spiral stellar arms ? *We expect that dynamo and MHD theories ought to make a **prediction on** their importance and implications.for spiral arm formation or maintenance.*

5. **Conclusion.**

Above we investigated large-scale galactic magnetism, observationally (Tables and Figures), and searched for predictions. What we found is these:

1) We showed a recent delineation of the location of the different arm tracers, including the peak of the total

synchrotron emission in a spiral arm offset by **about 220 pc from the hot dust lane** – see Figure 1. The position of the peak total synchrotron emission is within a spiral arm, located between the starformation region (orange lane in Fig.1), and the broad diffuse CO lane and Potential Minimum of the wave (blue lane in Fig.1); this interesting location should be a **prediction** of the density-wave theory and of a Magneto-Hydro-Dynamo theory.

2)    Quite a lot of bibliography was done in order to summarise the current knowledge related to an annulus with a reversed galactic magnetic field (see Table 1). After re-calibrating to a common distance of the Sun to the Galactic Center, we propose a statistical analysis to delimitate the reverse annulus (counterclockwise galactic magnetic field) – see Table 1 and Figure 2. **We find an annulus width of 2.1 kpc.** We expect that value to be **predicted** by a dynamo theory suitably fitted to the Milky Way conditions.

3)    By superimposing the magnetic field reversed annulus on a recent map of the spiral arm locations, looking for a match (yielding our useful Figure 2). **Thus the field-reversed annulus encompasses the Crux-Centaurus arm (in Galactic Quadrant IV) and the Sagittarius arm (in Galactic Quadrant I). These are arm segments, not full spiral arms.** One sees these as arm segments, not as full long arms. The author does not know of a galactic dynamo theory that mimics these magnetic field observations (clockwise and counterclockwise directions) in the same long spiral arm. The radial location of the reversed annulus (Table 1; Figure 2), led us to identify the location of the annulus with respect to the nearby spiral arms; existing theories (Galactic dynamo, MHD) should make such a **prediction**, including the effects on a spiral arm when encountering a Galactic-wide magnetic field reversal.

4)    Note that observations **do not** have a clockwise magnetic field in *all* spiral arms, nor a counterclockwise field in *all* interarms, contrary to preliminary numerical models (Fig.7 in Gomez & Cox 2002; Fig. 8 in Han 2017; Fig. 2 in Kong 2022).

5)    We made a look of six known magnetised supershells near the Sun (within 400 pc) – see Table 2 and Figure 3 with expanding supershells pushing around and compressing the regular large-scale disk magnetic field. This compression must also be occurring everywhere in our Galactic disk, around stellar supershells. What is the role of the interstellar turbulence, with respect to the Galactic-wide magnetic field? Dynamo and MHD theories ought to make a **prediction on** their magnetic implications.for spiral arm birth and evolution.

We made a bibliographic search for Galactic-wide dynamo theories (Section 3) predicting an annulus with a reversed magnetic field direction (but often too large as compared to our observations).

Density wave models can account for spiral arms in the Milky Way, including the separations of arm tracers in each spiral arm (Fig.1). Thus density waves should be incorporated in galactic dynamo theories (Rohde et al 1999; Chiba & Tosa 1990; or Chamandy et al 2013). We are still searching for the answers to four predictions (above): location of peak synchrotron inside the arm width, proper width and location of a reversed magnetic annulus, implication of magnetism for a typical spiral arm, magnetic effects of interstellar turbulence on the regular magnetic field (clockwise or counterclockwise).

**Acknowledgements.** The figure production made use of the PGPLOT software at the NRC Canada in Victoria.

**Data Availability:** All data underlying this article are available in the article, and / or will be shared on reasonable request to the Corresponding author.

**Conflicts of Interest:** The author declares no conflicts of interest regarding the publication of this paper.